\RequirePackage{fix-cm}
\documentclass[smallextended]{svjour3}       
\smartqed  
\usepackage{graphicx,color}
\newcommand{\cd}{\makebox[0.08cm]{$\cdot$}}
\begin{document}

\title{Equation for the Nakanishi weight function using the inverse  Stieltjes transform
}


\author{V.A.~Karmanov    \and J.~Carbonell  \and T.~Frederico}


\institute{V.A.~Karmanov \at
             Lebedev Physical Institute, Leninsky Prospekt 53, 119991 Moscow, Russia \\
              \email{karmanov@sci.lebedev.ru}           
           \and
           J.~Carbonell \at
             Institut de Physique NucleÌaire, Universit\'e Paris-Sud, IN2P3-CNRS, 91406 Orsay Cedex, France\\
              \email{carbonell@ipno.in2p3.fr} 
              \and 
              T.~Frederico \at
              Instituto Tecnol\'ogico de Aeron\'autica, DCTA,
12228-900, S. Jos\'e dos Campos,~Brazil\\
     \email{tobias@ita.br}
}

\date{Received: date / Accepted: date}

\maketitle

\begin{abstract}
The bound state Bethe-Salpeter amplitude was expressed  by Nakanishi 
in terms of  a smooth  weight function $g$. 
By using the generalized Stieltjes transform, we derive
an integral equation for the Nakanishi function $g$ for a bound state case. 
It has the standard form $g= \hat{\mathcal V} \, g$,  where $\hat{\mathcal V} $ is a two-dimensional integral operator. 
The prescription for obtaining the kernel ${\mathcal V} $  starting with the kernel $K$ of the Bethe-Salpeter equation is given. 
\end{abstract}

\section{Introduction}
\label{intro}

The Bethe-Salpeter (BS) approach \cite{SB_PR84_51} is a powerful tool to study the properties of the relativistic few-body systems. 
To avoid the problems related to  singularities of the BS amplitude in Minkowski space, the BS equation is often transformed, by means of the Wick rotation \cite{Wick}, in the Euclidean space. In this way, one finds the same spectrum as for the equation in the Minkowski space. However, corresponding Euclidean BS amplitude is not enough for calculating the observables.  As demonstrated in \cite{bs-long} in case of the ladder kernel -- which displays the strongest (pole) singularities --   it is possible to find  the bound  states and the scattering solutions in Minkowski space by a direct numerical calculations.
However, this problem  is much more complicated  to deal with  than in the Euclidean space. 

Another efficient approach to solve the BS  equation, proposed in \cite{KusPRD95} and developed in a series papers \cite{bs1,bs2,FrePRD14,FSVPRD,dPaPRD16,GutPLB16,nak,nak1}, relies on the Nakanishi representation \cite{nakanishi} of the BS amplitude   
$\Phi(k,p)$ in terms of a non-singular  weight function $g(\gamma,z)$. This representation reads:
\begin{equation}\label{eq1}
\Phi(k,p) =  \int_{0}^{\infty} d\gamma{'}  \int_{-1}^{1} dz^{'} \frac{g(\gamma^{'}, z^{'})}{\left( \gamma^{'} + \kappa^2  -{k}^2 -p\cd {k}\,z^{'}-i\epsilon \right)^3},\,\,\,
\end{equation}
where  $\kappa^2=m^2-\frac{1}{4}M^2,$
$m$ is the constituent mass and $M$ is the total mass of the bound state ($p^2=M^2$).
Once $g$ is known,  one can compute the BS  amplitude by means of (\ref{eq1})
and express through $g$ the observables, like  the electromagnetic form factors   \cite{CarEPJA09}.  
For the massless ladder exchange ($\mu=0$), the integral (\ref{eq1}) turns into a one-dimensional one (over $z'$ only). This 1D representation was used to solve the BS equation long ago, before inventing by Nakanishi the representation (\ref{eq1}), in the pioneering researches  by Wick \cite{Wick} and Cutkosky \cite{Cutkosky}.

In  \cite{bs1}, the following double integral equation for  the Nakanishi weight function $g$ was established:
\begin{equation} \label{bsnew}
\int_0^{\infty}\frac{g(\gamma',z)d\gamma'}{\left[\gamma+\gamma' +z^2 m^2+(1-z^2)\kappa^2\right]^2} =
\int_0^{\infty}d\gamma'\int_{-1}^{1}dz'\;V(\gamma,z;\gamma',z') g(\gamma',z'),
\end{equation}
(written symbolically as $\hat{L}\, g=\hat{V} \, g$)
where  $V$ is expressed via the BS kernel $K$. 
The solution  $g$ of the equation (\ref{bsnew}) was found  in the case of an OBE  ladder BS kernel  in \cite{bs1} and for the ladder plus cross-ladder one  in \cite{bs2,ff_cross}. 

Equation (\ref{bsnew}) contains integral terms in both sides, though the integral in the l.h.-side is one-dimensional.
This fact generates some instability of  its numerical  solution, 
since the discretizing of its l.h.-side results into an ill-conditioned matrix \cite{bs1}.
An equation for  $g$,  getting rid of the left-hand side integral term of (\ref{bsnew}), was first derived in  \cite{FrePRD14}. 
It has the following form:
\begin{equation}\label{g_Vg}
g(\gamma,z)=\int_0^{\infty}d\gamma'\int_{-1}^{1}dz'\;{\mathcal V}(\gamma,z;\gamma',z') g(\gamma',z').
\end{equation}
Like in (\ref{bsnew}), the r.h.-side of the equation (\ref{g_Vg}) contains the double integral.
This derivation was based on the uniqueness\footnote{The uniqueness means that if the integral (\ref{eq1}) is zero, then the function $g(\gamma,z)$ is also zero.}  of the Nakanishi representation (\ref{eq1}) and it was fulfilled for the ladder kernel only. 
The analytical expression for the kernel ${ \hat{\mathcal V}}$ in eq. (\ref{g_Vg}) is given in \cite{FrePRD14} and also below in Sec. \ref{deriv}.
The straightforward application of the derivation \cite{FrePRD14} to the same equation with a non-ladder kernel would meet some  algebraic difficulties. 

In a recent work \cite{nak2} another method of transforming
(\ref{bsnew})  in the canonical form (\ref{g_Vg}) was found. 
It does not rely on the uniqueness of the Nakanishi representation nor on the ladder kernel  
but  on using the explicit form of the inverse operator $\hat{L}^{-1}$ in  equation $\hat{L}\, g=\hat{V} \, g$  for an arbitrary kernel. 
It was indeed noticed that the integral in l.h.-side of (\ref{bsnew}) is the generalized Stieltjes transform whose inverse operator $\hat{L}^{-1}$ was derived in \cite{schw}.

We will review in this contribution   the main steps of \cite{nak2}  and give
simplified  expressions  for the kernel $\hat{\mathcal V}=\hat{L}^{-1}\hat{V}$ in (\ref{g_Vg}). In addition, we will study the limit of exchanged mass $\mu\to0$. 
We will show analytically that in this limit the equation (\ref{g_Vg}) turns into the Wick-Cutkosky equation \cite{Wick,Cutkosky}.

\section{Deriving the integral equation for $g$}\label{deriv}

The results that follow are based on the observation \cite{nak2}, that  the integral in l.h.-side of  (\ref{bsnew})  can be trivially related to the Stieltjes transform which is inverted analytically \cite{schw,Sumner_49}.
For the integral relation in the form, close to the l.h.-side of eq. (\ref{bsnew}):
\begin{equation} \label{L}
f(\gamma)\equiv  \int_0^{\infty} \;d\gamma' \;L(\gamma,\gamma') g(\gamma')   = \int_0^{\infty} \;d\gamma' \; \frac{g(\gamma')}{ (\gamma'+\gamma+b)^2},
\end{equation}
denoted symbolically as  $f = \hat{L} \;  g$, a straightforward  application of  the inverse Stieltjes transform  to  (\ref{L}) gives
\begin{equation} \label{L3}
g(\gamma)= \hat{L}^{-1} f=\frac{\gamma}{2\pi}\int_{-\pi}^{\pi}d\phi\, \;e^{i\phi}\; f(\gamma \; e^{i\phi}-b ).
\end{equation}

By applying the inverse integral transform (\ref{L3}) to both sides of the equation (\ref{bsnew}), we obtain
the equation for the Nakanishi weight function $g$ in the canonical form  (\ref{g_Vg}),
where 
\begin{equation} \label{L7}
{\mathcal V}(\gamma,z;\gamma',z') =\frac{\gamma}{2\pi}\int_{-\pi}^{\pi} \;d\phi\;  e^{i\phi} \; V\Bigl(\gamma e^{i\phi}-z^2 m^2-(1-z^2)\kappa^2,z;\gamma',z'\Bigr).
\end{equation}

The relation between the original kernel $K$  appearing in the BS equation and the  kernel $V$ in (\ref{bsnew})  and  (\ref{L7}) was derived in Ref. \cite{bs1}.
In  the case of the OBE  kernel
\begin{equation} \label{K}
K(k,k')= -{g^2 \over (k-k')^2-\mu^2+i \epsilon },   \qquad g^2= 16\pi m^2 \alpha
\end{equation}
$V$ takes the form  \cite{bs1}:
\begin{equation} \label{Kn}
V(\gamma,z;\gamma',z')=\left\{
\begin{array}{ll}
W(\gamma,z;\gamma',z'),&\mbox{if $-1\le z'\le z\le 1$}\\
W(\gamma,-z;\gamma',-  z'),&\mbox{if $-1\le z\le z'\le 1$}
\end{array}\right.
\end{equation}
where:
\begin{eqnarray}\label{W}
W(\gamma,z;\gamma',z') &=&
\frac{\alpha m^2}{2\pi}  \frac{(1-z)^2}{[\gamma+z^2m^2+(1-z^2)\kappa^2] b_2^2(b_+ -b_-)^3}  \\
&\times&\left[ \frac{(b_+ -b_-)(2b_+ b_- -b_+ -b_-)}{(1-b_+)(1-b_-)}  +  2b_+ b_- \log \frac{b_+ (1-b_-)}{b_- (1-b_+)}\right], 
\nonumber
\end{eqnarray}
\begin{eqnarray*}
\mbox{and}\quad b_0 &=& (1-z)\mu^2, \,\,\,\,\, b_\pm = -\frac{1}{2b_2} \;\left( b_1 \pm \sqrt{b_1^2-4b_0b_2}\right), \\
b_1 &=& \gamma+\gamma' - (1-z)\mu^2 - \gamma' z -\gamma z' +(1-z')\left[z^2m^2+(1-z^2)\kappa^2\right], 
\nonumber\\
b_2 &=& -\gamma (1-z')-(z-z') \left[  (1-z)(1-z')\kappa^2+(z+z'-zz') m^2 \right].
\nonumber
\end{eqnarray*}
The kernel ${\mathcal V}$ of  Eq. (\ref{g_Vg})  is determined by inserting  (\ref{Kn})  in the integral (\ref{L7}).  

As mentioned in the Introduction,  the canonical form (\ref{g_Vg}) of Eq. (\ref{bsnew}) was first derived   in \cite{FrePRD14} 
under the hypothesis of the ladder kernel and an expression for the kernel ${\mathcal V}$ was 
found.\footnote{Following \cite{Gianni},
a misprint in  the kernel sign in \cite{FrePRD14} was corrected here and in \cite{nak2}.}
We calculate analytically some integrals  and transform the kernel  \cite{FrePRD14}  to the form:
\begin{equation}\label{Vf}
{\mathcal V}(\gamma,z;\gamma',z') = + \frac{\alpha m^2}{2\pi}\times
\left\{
\begin{array}{ll}
\displaystyle{h(\gamma,-z;\gamma',-z')}, &\quad \mbox{if $-1\le z'\le z\le 1$} \\
\displaystyle{h(\gamma,z;\gamma',z')}, &\quad  \mbox{if $-1\le z\le z'\le 1$}
\end{array}
\right.
\end{equation}
The function $h$ is obtained from eqs. (27), (28) in  \cite{FrePRD14}:
\begin{equation}\label{h}
h(\gamma,z;\gamma',z')=Q(\gamma',z')+\theta(\eta)P(\gamma,z;\gamma',z').
\end{equation}
The first term $Q$ is given by
\begin{eqnarray}\label{Q}
Q(\gamma',z')&=& \int_0^{\infty}\chi(y)dy =-\frac{A'}{AA_s}
-\left\{
\begin{array}{ll}
\displaystyle{\frac{2\mu^2}{A_s^{3/2}}\left(2\arctan\frac{A'}{\sqrt{A_s}}-\pi\right)}, &
\mbox{if $A_s>0$}
\\
\displaystyle{\frac{2\mu^2}{\mid A_s\mid^{3/2}}\log\frac{A'+\sqrt{ \mid A_s\mid }}{A'-\sqrt{\mid A_s\mid }}},&
\mbox{if $A_s<0$}
\end{array}\right.
\nonumber
\end{eqnarray}
where  the integrand is given by\footnote{A misprint in eq. (28) of \cite{nak2} for $\chi(y)$ was corrected in (\ref{chi}):
$$ \chi(y)=\frac{y^2}{(A+y^2+A'y+\mu^2)^2}\to \chi(y)=\frac{y^2}{(Ay^2+A'y+\mu^2)^2}. $$}
\begin{equation}\label{chi}
\chi(y)=\frac{y^2}{(Ay^2+A'y+\mu^2)^2}
\end{equation}
and $A_s=4A\mu^2-{A'}^2$, $A=\gamma'+m^2-\frac{1}{4}(1-{z'}^2)M^2 $, $A'=\gamma'+\mu^2$

The argument $\eta$ of the theta-function in the second term in (\ref{h}) reads
\begin{equation}\label{eta}
\eta=\gamma\frac{1+z'}{1+z}-\mu^2-\gamma'-2\mu\sqrt{\gamma'+m^2-\frac{1}{4}(1-{z'}^2)M^2}=-B-2\mu\sqrt{A}.
\end{equation}
The function $P(\gamma,z;\gamma',z')$  in (\ref{h}) has the form:
\begin{equation}\label{P}
P(\gamma,z;\gamma',z')=\frac{B}{\gamma A \Delta}\frac{1+z}{1+z'}-C
\end{equation}
where
\[
B=-\gamma\frac{1+z'}{1+z}+\mu^2+\gamma',\quad \Delta=\sqrt{B^2-4\mu^2A} \]
\[C=\int_{y_-}^{y_+}\chi(y)dy=\hat{\chi}(y_+)-\hat{\chi}(y_-),\quad y_{\pm}=\frac{-B\pm\Delta}{2A} \]
\begin{eqnarray}\label{hatchi}
\hat{\chi}(y)&=&  \frac{A'\mu^2-2A\mu^2 y+{A'}^2y}{AA_s[\mu^2+y(A'+Ay)]}
+\left\{
\begin{array}{ll}
\displaystyle{\frac{4\mu^2}{A_s^{3/2}}\arctan\frac{A'+2Ay}{\sqrt{A_s}}},   &\quad\mbox{if $A_s>0$}
\\
\displaystyle{\frac{2\mu^2}{  \mid A_s\mid^{3/2}}
\log\frac{  {A'+2Ay\over\sqrt{\mid A_s\mid}} +1}{ {A'+2Ay\over\sqrt{\mid A_s\mid}}-1}},   &\quad\mbox{if $A_s<0$}
\end{array}\right.
\nonumber
\end{eqnarray}

Kernel ${\cal V}$ is determined above by two quite different expressions: eq. (\ref{L7}) found in \cite{nak2} and  eq. (\ref{Vf}) found in \cite{FrePRD14}. 
It has not been possible to prove  their identity analytically but  only by their numerical comparison  \cite{nak2}.
It is worth noticing however that, contrary to $V$,  kernel  ${\cal V}$ displays some singularities in variable $\gamma$ and requires
some careful treatment. 

The coupling constants $\alpha$ and the Nakanishi functions $g(\gamma,z)$ for selected values of $M$ and $\mu$ found in \cite{FrePRD14} from  equations (\ref{bsnew}) and (\ref{g_Vg})   coincide with each other within numerical accuracy as well as with the Euclidean BS results.  

\section{The limit $\mu\to 0$}
In the $\mu=0$ case, which constitutes the original Wick-Cutkosky model \mbox{\cite{Wick,Cutkosky}}, 
kernel (\ref{Kn}) obtains a simple analytical form:
\begin{eqnarray}\label{KK}
V(\gamma,z;\gamma',z')&=&\frac{\alpha m^2}{2\pi}
\frac{1}{\Bigl[\gamma+z^2m^2+(1-z^2)\kappa^2 \Bigr]\Bigl[\gamma'+{z'}^2m^2+(1-{z'}^2)\kappa^2 \Bigr]}
\nonumber\\
&\times&\left\{
\begin{array}{ll}
\displaystyle{\frac{1}{\Bigl[\gamma +\gamma'\frac{(1+z)}{(1+z')}
+z^2m^2+(1-z^2)\kappa^2\Bigr]}
\frac{(1+z)}{(1+z')}},&\quad\mbox{if $z<z'$}
\\
\displaystyle{\frac{1}{\Bigl[\gamma +\gamma'\frac{(1-z)}{(1-z')}
+z^2m^2+(1-z^2)\kappa^2\Bigr]}
\frac{(1-z)}{(1-z')}},&\quad\mbox{if $z>z'$}
\end{array}
\right.
\end{eqnarray}
Inserting this expression in (\ref{L7}) and  integrating  over $\phi$ one gets for ${\cal V}$:
\begin{equation}\label{Nu0}
{\mathcal V}(\gamma,z;\gamma',z')=\frac{\alpha m^2}{2\pi\gamma'}
\displaystyle{\frac{1}
{\left[\gamma' +m^2-\frac{1}{4}(1-{z'}^2)M^2\right]}}
\times
\left\{
\begin{array}{ll}
\theta\left(\gamma'\frac{1+z}{1+z'}-\gamma\right),&\;\mbox{if $z<z'$}
\\
\theta\left(\gamma'\frac{1-z}{1-z'}-\gamma\right),&\;\mbox{if $z>z'$}
\end{array}
\right.
\end{equation}
Following  \cite{bs1}, we look for a solution of (\ref{g_Vg}) in the form 
$$
\bar{g}(\gamma,z)=\delta(\gamma)g(z).
$$
This form is justified provided the equation for $g(z)$ does not depend on $\gamma$. Substituting the former expression  in (\ref{g_Vg}), we obtain:
 \begin{equation}\label{g0a}
\delta(\gamma)g(z)=\int_0^{\infty} d\gamma' \int_{-1}^1dz'{\cal V}(\gamma,z;\gamma',z')\delta(\gamma')g(z')
\end{equation}
Integrating\footnote{On the one hand, the integrand in (\ref{g0a}) contains the expression $\frac{1}{\gamma'}\delta(\gamma')$ which is infinite. 
On the other hand, the integration of r.h.-side of (\ref{g0a}) over $\gamma$ gives, due to the theta-function, the factor $\gamma'\frac{1\pm z}{1\pm z'}$.  
$\gamma'$ from this factor cancels $\gamma'$ in the denominator. In this way, we obtain a finite and certain result for the product $\infty\cdot 0$.}
both sides of (\ref{g0a}) over $\gamma$ we find:
\begin{equation}\label{g0b}
g(z)=\int_{-1}^1d z' \widetilde{V}(z,z')g(z'),
\end{equation}
with
\begin{equation}\label{tV}
\widetilde{V}(z,z')=\frac{\alpha m^2}{2\pi} 
\frac{1}{\left[m^2-\frac{1}{4}(1-{z'}^2)M^2\right]}
\left\{
\begin{array}{ll}
\displaystyle{\frac{1+z}{1+z'}},&\quad\mbox{if $z<z'$}
\\
&
\\
\displaystyle{\frac{1-z}{1-z'}},&\quad\mbox{if $z<z'$}
\end{array}
\right.
\end{equation}
which exactly coincides with the Wick-Cutkosky equation  \cite{Wick,Cutkosky}. 
Notice that the same limit $\mu\to 0$ can be  derived starting from (\ref{g_Vg}) and  kernel (\ref{Vf}).

\section{Conclusion}\label{concl}

Starting with an equation for the Nakanishi weight function $g$ of the Bethe-Salpeter amplitude in the form $\hat{L}\,g=\hat{V}\,g$, 
previously established in \cite{bs1,bs2}, and  using an analytic inversion of the Stieltjes integral transform,
we transformed this equation in the canonical form $g=\hat{\mathcal V}g$.  

This result generalizes, to an arbitrary  kernel given by a set of  irreducible Feynman graphs, the equation found in \cite{FrePRD14} 
which was stablished  for a ladder kernel.
This generalization strongly expands the applicability of the Nakanishi representation to find the solution of the Bethe-Salpeter equation, 
as well as its applications. 
Equation (\ref{g_Vg}) can be extended straightforwardly to the two-fermion system starting, for example, from \cite{dPaPRD16}.

The possibility to apply an analytic inversion of the Stieltjes transform can be useful in other fields
of nuclear and hadronic physics, where the use of  integral transforms was  pioneered 
by  \cite{Efros_SJNP_1985,Efros:2007nq,Giussepina_EFB23_2016},
in order to avoid the instabilities of the numerical inversion.

\section*{Aknowledgements}
We are indebted to G. Salm\'e for useful discussions. T.F. thanks CNPq, 
CAPES and FAPESP of Brazil. V.A.K. thanks the support of FAPESP, the
grant \#2015/22701-6 and is sincerely grateful for kind hospitality to 
Theoretical Nuclear Physics Group in ITA, S\~{a}o Jos\'e dos Campos,  
Brazil, where the main part of this research was carried out.


\end{document}